\documentclass[aps,twocolumn,showpacs]{revtex4}
\usepackage{graphics,graphicx}
\usepackage{color}
\usepackage{wrapfig}
\usepackage{enumerate}
\usepackage{amssymb,amsmath}
\usepackage{bm}

\usepackage[normalem]{ulem} 

\begin{document}
\unitlength = 1mm
\preprint{LA-UR-12-26093}
\title{
	Multi-discontinuity algorithm for world-line Monte Carlo simulations
}

\author{Yasuyuki~Kato}
\email{ykato@lanl.gov}
\affiliation{Theoretical Division, T-4 and CNLS, Los Alamos National Laboratory, Los Alamos, NM 87545}

\date{\today}
%
\begin{abstract}
We introduce a multi-discontinuity algorithm for the efficient global update of world-line configurations in Monte Carlo simulations of interacting quantum systems.
This algorithm is a generalization of the two-discontinuity algorithms introduced in Refs.
[N.~Prokof'ev, B.~Svistunov, and I.~Tupitsyn, Phys. Lett. A {\bf 238}, 253 (1998)] 
and [O.~Sylju{\aa}sen and A.~Sandvik, Phys. Rev. E {\bf 66}, 046701 (2002)] .
This generalization is particularly effective for studying Bose-Einstein condensates (BEC) of composite particles.
In particular, we demonstrate the utility of the generalized algorithm by simulating a Hamiltonian for an $S=1$ anti-ferromagnet with strong uniaxial single-ion anisotropy. The multi-discontinuity algorithm not only solves the freezing problem 
that arises in this limit, but also allows the efficient computing of 
the off-diagonal correlator that characterizes a BEC of composite particles.
\end{abstract}
\maketitle
{\it Introduction}:
Cooperative phenomena in quantum systems (e.g., superfluidity, Bose-Einstein condensation, superconductivity, magnetism, etc.) 
have fascinated physicists for decades.
The study of these phenomena often requires one to consider interaction regimes
for which traditional perturbative approaches do not work.
World-line quantum Monte-Carlo (QMC) simulations based on the Feynman path integral formulation 
have demonstrated to be a very powerful method for studying systems in these regimes \cite{ceperley1995,kawashima2004}.
The main reasons are that the QMC method is unbiased, and its results are exact within statistical error.
Therefore, it is imperative to expand the range of applicability of this method  by developing QMC algorithms for
problems that can not be simulated with state-of-the-art numerical techniques.

Although the worm algorithm \cite{prokof1998} and the directed-loop algorithm (DLA) \cite{syljuasen2002},
are the most efficient strategies for global updates in the QMC simulations,
they have difficulties when applied to systems exhibiting Bose-Einstein condensates (BECs) of composite particles.
BECs of composite particles appear in different contexts, 
such as bosonic gases of two species ($a$ and $b$) of ultra cold atoms \cite{kuklov2003,soyler2009} and
homonuclear bosonic gases ($^{23}{\rm Na}_2$ \cite{xu2003}, $^{133}{\rm Cs}_2$ \cite{herbig2003}, etc.).
Another simple example of BECs of composite particles is the ferro-nematic magnetic ordering
that appears in quantum magnets with single-ion anisotropy \cite{wierschem2012}.
In general, high spin systems ($S\geq 1$) whose total magnetization along a particular axis is conserved (it will be our $z$-axis) 
can exhibit a local order parameter of the form
$\langle (S^{+})^n \rangle$ ($2\leq n \leq 2S$) in the absence of  magnetic and any lower moments, i.e., $\langle (S^{+})^m \rangle=0$
for $m<n$. These order parameters correspond to different moments of a distribution of electric currents. For instance,
$n=2$ corresponds to quadrupolar or nematic order, and $n=3$ corresponds to octupolar order.
After mapping the spin system into a gas of bosons, the phase with order parameter $\langle (S^{+})^n \rangle \neq 0$ is mapped into a BEC of 
composite particles containing $n$-bosons. For instance, nematic ordering is mapped into a BEC of pairs of bosons \cite{wierschem2012}. 

QMC simulations sample and update the $d+1$ dimensional world-line configurations of the quantum system
under consideration ($d$ is the spatial dimension).
Standard algorithms insert a pair of discontinuities
{\it head} and {\it tail} in the world-line configuration. 
The head propagates until it meets the tail \cite{kawashima2004}.
The head changes the local states of the world-line configuration along its way.
As a result, states on the loop drawn by the head's trajectory are globally updated.
For bosonic systems, the discontinuities typically correspond to the creation, $a^{\dag}$, 
or annihilation, $a$, of a particle.
For QMC simulations of mixtures of bosonic gases 
\cite{kuklov2003,kuklov2004,soyler2009,guglielmino2010,ohgoe2011,capogrosso2011,capogrosso2011b,pollet2012},
including $a$-type and $b$-type bosons, the discontinuities can 
be of the form $a^{\dag}b$ or $b^{\dag}a$ that exchange the type of boson \cite{ohgoe2011}. 
However, we will see that two discontinuities are not enough to perform efficient simulations in certain cases
because of the so-called ``freezing problem". 
To solve this problem, algorithms with two or more pairs of discontinuities corresponding to $a$, $a^\dag$, $b$ or $b^\dag$ 
have been considered as a generalization of the worm algorithm
(e.g. Refs.~\cite{soyler2009,capogrosso2011b,pollet2012}).
Furthermore the two-discontinuity algorithms often suffer from a serious slowing down problem
for calculations of off-diagonal correlators like 
$\langle a_{\bm r}^{\dag}(\tau)b_{\bm r}(\tau) b_{\bm r'}^{\dag}(\tau')a_{\bm r'}(\tau') \rangle$,
which characterize the condensation of composite particles.

In this paper, we present a method called the multi-discontinuity algorithm (MDA) that incorporates an arbitrary number
of discontinuities in order  to solve the freezing problem.
We will show that the MDA can be used for simulating BECs of composite particles 
that cannot be simulated efficiently with two-discontinuity algorithms.
Although the MDA can be applied  to many different situations,
we will consider a simple example of an $S=1$ antiferromagnetic ($J>0)$ Heisenberg model 
with external magnetic field, $B$, and single-ion uniaxial anisotropy $D$.
We will see that the MDA successfully  explores the ferronematic phase  of this model, while the DLA fails because of the freezing problem. 
Since one additional discontinuity is enough for simulating the $S=1$ system, we restrict to three discontinuities in this paper.
The extension to the algorithm to more than three discontinuities is straightforward.

{\it Model}:
We consider the spin-one Hamiltonian
\begin{eqnarray}
\mathcal{H} &=& J\sum_{\langle {\bm r}, {\bm r}'\rangle} {\bm S}_{\bm r} \cdot {\bm S}_{\bm r'}
+D\sum_{\bm r} {S^{z}_{\bm r}}^2-B\sum_{\bm r} {S^{z}_{\bm r}},
\label{eq:hamiltonian0}
\end{eqnarray}
where the summation of ${\langle {\bm r}, {\bm r}'\rangle}$ runs over all the nearest-neighbor bonds and the model is defined on
$d$-dimensional hypercubic lattices \cite{hamer2010,wierschem2012}.
We restrict the model to the case of easy-axis single-ion anisotropy  $D<0$.
In the absence of a magnetic field ($B=0$), the ground state of ${\mathcal H}$ exhibits Ising-like antiferromagnetic ordering along the $z$ axis (AFM-z).
This AFM-z phase is destroyed by the application of a strong enough magnetic field $B$ via a
first-order quantum phase transition to a new phase that exhibits  ferronematic (FN) ordering \cite{wierschem2012}.
A further increase of the magnetic field leads to a second order phase transition to the fully polarized phase.  
As we will see later, the MDA is necessary for finding and studying the FN phase that is characterized by the FN susceptibility,
\begin{eqnarray}
\chi_{\rm FN} &=& \frac{1}{L^d}
\sum_{{\bm r}, {\bm r}'} \int_{0}^{\beta} d\tau 
\left\langle
S^{+}_{\bm r}(\tau)S^{+}_{\bm r}(\tau)S^{-}_{{\bm r}'}(0)S^{-}_{{\bm r}'}(0)\right\rangle.
\label{chi}
\nonumber\end{eqnarray}

\begin{figure}[htpb]
\includegraphics[angle=0,width=6cm,trim= 80 60 120 40,clip]{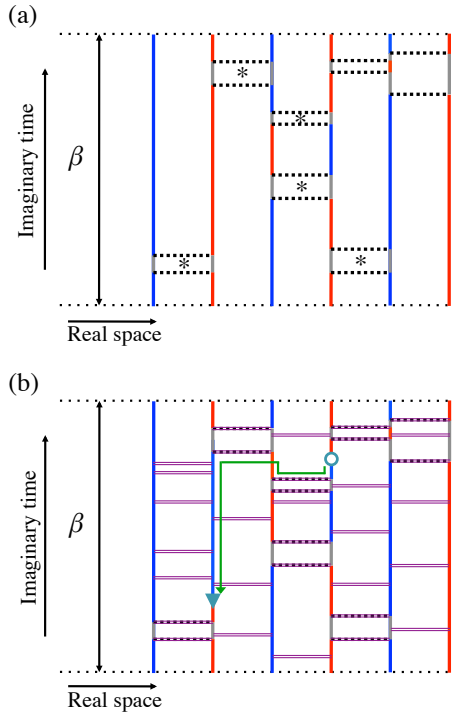}
\caption{
	(Color online) (a) Typical world-line configuration for a chain of six sites ($d=1$),  a large $|D|\gg J$ ratio, and very low temperature 
	$\beta J \gg 1$. Red, gray, and blue solid lines correspond to $S^z=-1,0$, and $+1$, respectively.
	The stars ($*$) mark all the bubbles that represent quantum fluctuations. 
	(b) A world-line configuration with a pair of discontinuities called a worm in the directed-loop algorithm. 
	The purple horizontal double lines are the scattering objects (vertices) for the discontinuity.
	}
\label{fig:fig1}
\end{figure}

{\it Algorithm}:
World-line configurations in the $d+1$ dimensional space introduced in the Feynman path integral formulation are sampled by the QMC method \cite{kawashima2004}. 
Here we explain the MDA as a generalization of the DLA by using model \eqref{eq:hamiltonian0} as an example of its application. 
When the DLA is simply applied to model \eqref{eq:hamiltonian0} for $J>0$, 
it suffers from the negative sign problem.
To avoid the negative sign problem, we introduce a spin rotation by $\pi$ along the $z$ axis 
($S^{x,y}_{\bm r} \rightarrow -S^{x,y}_{\bm r}$) for the $B$ sublattice.
The expression of the Hamiltonian in the new basis is 
\begin{eqnarray}
\mathcal{H} &=& -J\sum_{\langle {\bm r}, {\bm r}'\rangle} 
\left(
     S^x_{\bm r} S^x_{\bm r'}
     +S^y_{\bm r} S^y_{\bm r'}
     -S^z_{\bm r} S^z_{\bm r'}
\right)\nonumber\\
&&+D\sum_{\bm r} {S^{z}_{\bm r}}^2-B\sum_{\bm r} {S^{z}_{\bm r}}.
\label{eq:hamiltonian}
\end{eqnarray}
We will start by applying the DLA to see how the freezing problem arises for $|D| \gg J$ and $B \gtrsim J$. 
Figure \ref{fig:fig1}(a) shows a typical world-line configuration appearing at $|D| \gg J$ and low temperature. 
Because of the strong anisotropy, $|D| \gg J$, the states $|S^z=\pm 1\rangle $ are the vast majority, while the $|S^z=0\rangle$ states rarely appear in the world-line configurations.  The density of these minority states is $\sim J^2/D$ (note that the volume of the $d+1$ dimensional space  has units of inverse  energy so the density of states has units of energy). 
Moreover, the $|S^z=0\rangle$ states typically appear forming bubbles that correspond to quantum fluctuations.
These quantum fluctuations are crucial for the stabilization of the FN ordering \cite{wierschem2012}. 
We will see that the bubbles of $|S^z=0\rangle$ states are the main source of difficulty for the DLA. 

\begin{figure}[htpb]
\includegraphics[angle=0,width=8.5cm,trim= 0 350 0 0,clip]{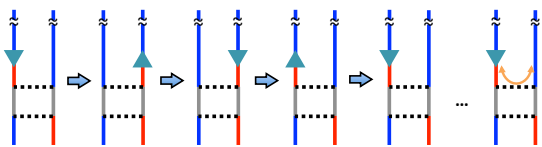}
\caption{
	(Color online)
	Local trapping of an $(S^+)^2$ discontinuity at a bubble. The vertices are not illustrated here for simplicity.
	}
\label{fig:fig2}
\end{figure}

In the DLA algorithm, the world-line configurations are globally updated by introducing a pair of discontinuities called {\it worm} and their scattering objects called {\it vertices}. The world-line configurations are updated according to the following protocol:
\begin{description}
	\item[Step 1] Place vertices stochastically (purple double line in Fig.~\ref{fig:fig1}(b)).
	The density of vertices at $( \langle {\bm r}, {\bm r}'\rangle, \tau)$ is a function of $(S^z_{\bm r} (\tau),S^z_{{\bm r}'} (\tau) )$.
	\item[Step 2] Choose a point using a uniform random distribution, create a pair of discontinuities (worm), 
	and assign one discontinuity to be the head (triangle object in Fig.~\ref{fig:fig1}(b)) and the other to be the  tail (circle object in Fig.~\ref{fig:fig1}(b)).
	\item[Step 3] Move the head until it hits a vertex or the tail. If it hits the tail, go to Step 3-1, otherwise, Step 3-2.
	\item[Step 3-1] Let the worm annihilate, and go back to Step 2, or go to Step 4. 
	\item[Step 3-2] Choose its new direction stochastically, and go to Step 3.  
	\item[Step 4] Erase all vertices after creating and annihilating the worm several times.
\end{description}
More details of the DLA can be found in Ref.~\cite{syljuasen2002,kawashima2004}.
The worm is generated by a  ``source  term'', $-\eta_0 \mathcal{Q}$, that  is added to the Hamiltonian  ($\eta_0>0$).
For the case of our model \eqref{eq:hamiltonian}, we choose $\mathcal{Q}$ as
\begin{eqnarray}
\mathcal{Q} &=& \sum_{\bm r} \left[ \frac{S^+_{\bm r}+ S^-_{\bm r} }{\sqrt{2}} 
 +\frac{(S^+_{\bm r})^2 + (S^-_{\bm r})^2}{2}  \right].
\end{eqnarray}
The discontinuity is created by any of the four terms $S^+_{\bm r}$, $S^-_{\bm r}$, $(S^+_{\bm r})^2$, or $(S^-_{\bm r})^2$.
We sample world-line configurations whose Boltzmann weights are ${\cal O}(\eta_0^0)$ and ${\cal O}(\eta_0^2)$. 
In other words, the allowed world-line configurations have only one warm or no discontinuity at all.
Because of the external field $B \gtrsim J$, the head created by $(S^+)^2$ is trapped locally in a bubble, as shown in Fig.~\ref{fig:fig2}.
The probability of escaping from such a trap is exponentially small in the number of vertices corresponding the $B$ term on the way to escape, 
and it becomes negligibly small 
because  the typical number of the vertices is ${\cal O}(D B/J^2) \gg {\cal O}(1)$. 
As a result, it takes an exponentially long time to complete a cycle of creation and annihilation of a worm.
This trapping problem makes the DLA inefficient for  simulating the FN phase of  ${\cal H}$.


\begin{figure}[htpb]
\includegraphics[angle=0,width=8.5cm,trim= 0 300 70 0,clip]{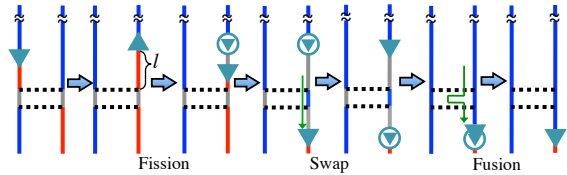}
\caption{
	(Color online)
	Example of a sequence of processes involving ``fusion", ``fission", and ``swap",  that allows the head to escape from the trap
	in the multi-discontinuity algorithm.
	}
\label{fig:fig3}
\end{figure}

The basic difficulty of the the DLA can be eliminated by considering an enlarged Hilbert space via the inclusion of a third discontinuity.
We add a new source term, $-\eta_1 {\mathcal Q}$ to ${\mathcal H}-\eta_0 {\mathcal Q}$, to generate a  third discontinuity that has a 
weight $\eta_1$, which can be different from $\eta_0$, and sample the world-line configurations whose weights are $O(\eta_0^0\eta_1^0)$, $O(\eta_0^2\eta_1^0)$, or $O(\eta_0^2\eta_1^1)$.
Accordingly, we consider three new processes that will be called ``{fusion}", ``{fission}" and ``{swap}" (see Fig.~\ref{fig:fig3} ).
These  new processes are executed  when the head stops at a newly introduced one-body vertex.
Although this new vertex is supposed to be created with a uniform distribution, in practice 
we create a virtual new vertex \cite{kato2007,kato2009} to avoid the unnecessary proliferation of vertices and the introduction of new free parameters.
Fission corresponds to the process by which one discontinuity  decays into two discontinuities.
We represent these processes by indicating the source  operators that create the initial and the final states:
[$(S^+)^2 \rightarrow S^+$ and $S^+$],
[$(S^-)^2 \rightarrow S^- $  and $ S^-$], 
[$S^+      \rightarrow (S^+)^2$ and $S^-$], and 
[$S^-       \rightarrow (S^-)^2$ and $S^+$]. 
The inverse processes correspond to fusions because they combine two discontinuities into a single one. 
The swap corresponds to the process of exchanging the head and one of the two tails.  
To include these three new processes, we need to replace Steps 3, 3-1, and 3-2 of the conventional DLA by the following steps: 
\begin{description}
	\item[Step 3] Calculate $l = - \ln(R)/(2\eta_1)$. $R$ is a random number with uniform distribution in the interval $(0,1]$. 
	(See the Appendix for details of virtual placement of vertices.)
	If $l \geq \tau_{0}$, where $\tau_{0}$ is the time interval  between the head and the next object the head finds on its way
	(a vertex or another discontinuity), go to Step 3-1, otherwise go to Step 3-2.
	\item[Step 3-1] Let the head go to the next object. If the next object is a vertex, go to Step 3-1-1, otherwise go to Step 3-1-2.
	\item[Step 3-1-1]  Choose a new direction of the head stochastically, and go back to Step 3. 
	\item[Step 3-1-2] If the total number of discontinuities  is $n_D=2$, go to Step 3-1-2-1, otherwise go to Step 3-1-2-2.
	\item[Step 3-1-2-1]  Let the discontinuities  annihilate each other, and go back to Step 2, or go to Step 4. 
	\item[Step 3-1-2-2] Let the two discontinuities fuse and go to Step 3. 
	\item[Step 3-2] Move the discontinuity through the time interval $l$. If $n_D=2$, go to Step 3-2-1, otherwise ($n_D=3$) go to Step 3-2-2.
	\item[Step 3-2-1] Let the head fission, and go to Step 3.
	\item[Step 3-2-2] Let the head freeze, exchange discontinuities as a new head to move randomly (swap), and go to Step 3.
\end{description}
The value of the free parameter $\eta_1$ is fixed so that 
the average path lengths of  the head with and without the third discontinuity are roughly the same. 

\begin{figure}[htpb]
\includegraphics[angle=0,width=8.5cm,trim= 40 50 220 480,clip]{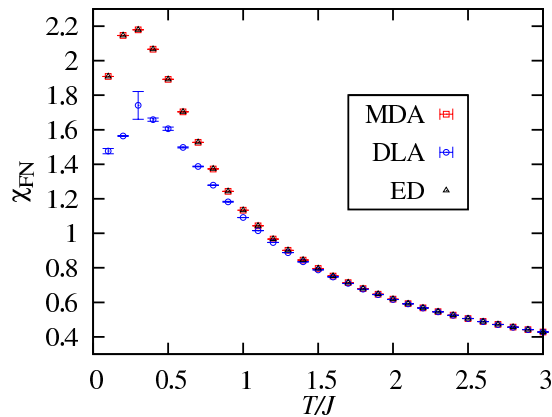}
\caption{
	(Color online)
	Ferronematic susceptibility, $\chi_{\rm FN}$, of model \eqref{eq:hamiltonian} defined on a ring $(d=1)$ of  $L=5$ spins. The Hamiltonian     
	parameters are $D/J=0$, and $B/J=0.3$. The three curves are the results obtained with
	the multi-discontinuity algorithm (MDA), the original directed-loop algorithm (DLA) and exact diagonalization (ED). 
	We performed $10^7$ Monte-Carlo sweeps for both MDA and DLA.
}
\label{fig:xmxx}
\end{figure}

{\it Example of application}:
Figure \ref{fig:xmxx} shows the FN susceptibility $\chi_{\rm FN}$ of model \eqref{eq:hamiltonian} defined on a ring ($d=1$) of  $L=5$ spins
and for Hamiltonian parameters $D/J=0$, and $B/J=0.3$. The three curves are the results obtained by using MDA, DLA and exact diagonalization (ED). 
The agreement between MDA and ED results is excellent. 
However, there is a rather large deviation between the  DLA and ED  results, although the agreement is very good for other physical quantities (e.g., energy and uniform magnetization) that are not shown.
This deviation confirms that 
DLA does not obtain an accurate estimate of the off-diagonal pair-pair correlation function 
$\left\langle S^{+}_{{\bm r}_+}(\tau_+)S^{+}_{{\bm r}_+}(\tau_+)S^{-}_{{\bm r}_-}(\tau_-)S^{-}_{{\bm r}_-}(\tau_-)\right\rangle$,
that determines the FN susceptibility because of a serious slowing down problem.
This correlation function is obtained by counting the number of events in which
only two discontinuities corresponding to $(S^+)^2$ and $(S^-)^2$ exist in whole space, with 
the $(S^+)^2$ discontinuity located at $({\bm r},\tau)=({\bm r}_+,\tau_+)$ and the $(S^-)^2$ discontinuity located at $({\bm r},\tau)=({\bm r}_-,\tau_-)$
\cite{ohgoe2011}. 
As it is clear from Eq.~\eqref{chi}, $\chi_{\rm FN}$ is the integration of the sum of this correlation function over all the possible values of the coordinates $({\bm r}_+,\tau_+)$ and  $({\bm r}_-,\tau_-)$.
We sample the whole length of the trajectory of the $(S^+)^2$ and $(S^-)^2$ discontinuities traveling when no other discontinuity exists.
As we explained in the previous section, the DLA suffers from a serious slowing down problem because the head of a worm made of a pair of  
$(S^+)^2$ and $(S^-)^2$ discontinuities cannot go through a bubble of $S^z=0$ states.
This limitation precludes the correct estimation of $\chi_{\rm FN}$, which is the crucial physical quantity for identifying the FN phase
in a finite size system.

\begin{figure}[htpb]
	\includegraphics[angle=0,width=8.5cm,trim= 40 50 220 480,clip]{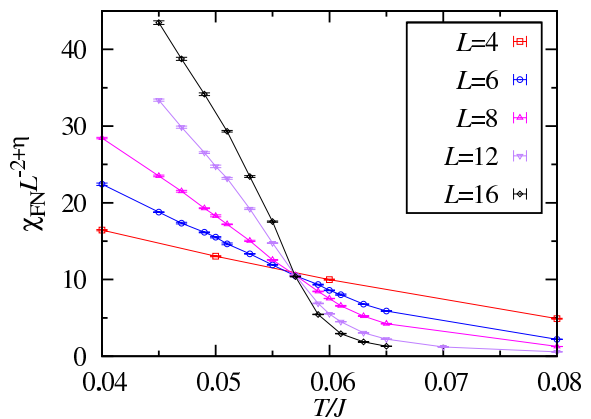}
\caption{
	(Color online)
	Finite-size scaling of ferronematic susceptibility for ${\cal H}$ defined on a simple cubic lattice $(d=3)$ with periodic boundary conditions.
	The Hamiltonian parameters are  at $D/J=-8$, and $B/J=6.25$.
	The estimated critical temperature is  $T_c=0.057$.
	We use the critical exponent $\eta = 0.038$ for the 3D XY universality class
	\cite{hasenbusch1999}.
	}
\label{fig:xmxxT}
\end{figure}

Figure \ref{fig:xmxxT} shows the temperature dependence of the conveniently scaled FN susceptibility, $\chi_{\rm FN}L^{-2+\eta}$, for the case of a simple cubic lattice and Hamiltonian parameters $D/J=-8$ and $B/J=6.25$. 
The common crossing point among the curves obtained with different system sizes indicates that there is a  finite temperature transition between the low temperature FN phase and a high temperature paramagnetic phase.
The order parameter of the FN phase is $\langle (S^+_{\bm r} )^2 \rangle$, which in bosonic language can be interpreted as a Bose-Einstein condensate of pairs of bosons. Thus, if continuous, the phase transition  should belong to the 3D XY universality class.
This is indeed the case according to the scaling analysis shown in Fig.~\ref{fig:xmxxT}: there is a common crossing point when we choose  the value of $\eta = 0.038$ that corresponds to the 3D XY universality class.

{\it Conclusions}:
In summary, we have introduced a MDA that overcomes some limitations of the two-discontinuity algorithms.   
The MDA helps to solve the  freezing problem that often arises in the simulation of 
strongly interacting quantum systems. The example that we discussed is just one out of many. 
Moreover the two-discontinuity algorithms do not satisfy ergodicity for a large class of physically relevant models.
In other words, there are physically possible world-line configurations that have not been sampled.
For instance, the MDA is also necessary for simulating  any quantum spin system with cubic single-ion anisotropy \cite{kawashima2004}.  
This is a quite common case for several Mott insulating materials such as  EuTiO$_3$ \cite{petrovic2012}.

The extension of the three-discontinuity algorithm  to $N$-discontinuity is straightforward, and it is
necessary for describing certain physical situations such as  $S\geq2$ antiferromagnets with cubic single-ion anisotropy. 
The only qualitative difference for $N >3$ is the possibility of creating or annihilating a new pair of discontinuities in world-line configurations that already contain 
a finite number of discontinuities. 
  For efficient simulations with $N>3$, we choose the swap or the fission stochastically when the head stops at a virtual vertex with $n_D<N$.
  (We choose the fission with probability 1 when $n_D=2$ and $N=3$.)
Although we introduced the MDA as a generalization of the DLA in this paper,
it is straightforward to modify the MDA in such a way that it becomes a generalization of the worm algorithm. 


The successful application of the MDA to the $S=1$ Heisenberg antiferromagnet with uniaxial single-ion anisotropy and Zeeman coupling to an external magnetic field demonstrates the relevance of this method.
We showed that the pair-condensation or ferronematic susceptibility, $\chi_{\rm FN}$, obtained from simulations based on the MDA 
accurately reproduces the ED results. The same is not true for the DLA.
Our MDA simulations also reproduce the expected  finite temperature second order phase transition between the paramagnetic and FN phases in a simple cubic lattice. 

\begin{acknowledgements}
I thank K.~Wierschem, P.~Sengupta, C.~D.~Batista, and N.~Kawashima for fruitful discussions. 
I also thank A.~Kuklov and T.~Ohgoe for useful comments on a draft.
This research was performed under the U.S. DOE Contract No.~DE-AC52-06NA25396 through the LDRD program,
and used resources of the NERSC, which is supported by the Office of Science of the U.S. DOE Contract No.~DE-AC02-05CH11231.
\end{acknowledgements}

\appendix*
\section{Fission and fusion}
\begin{figure}[htpb]
	\includegraphics[angle=0,width=8.5cm,trim= 150 40 150 30,clip]{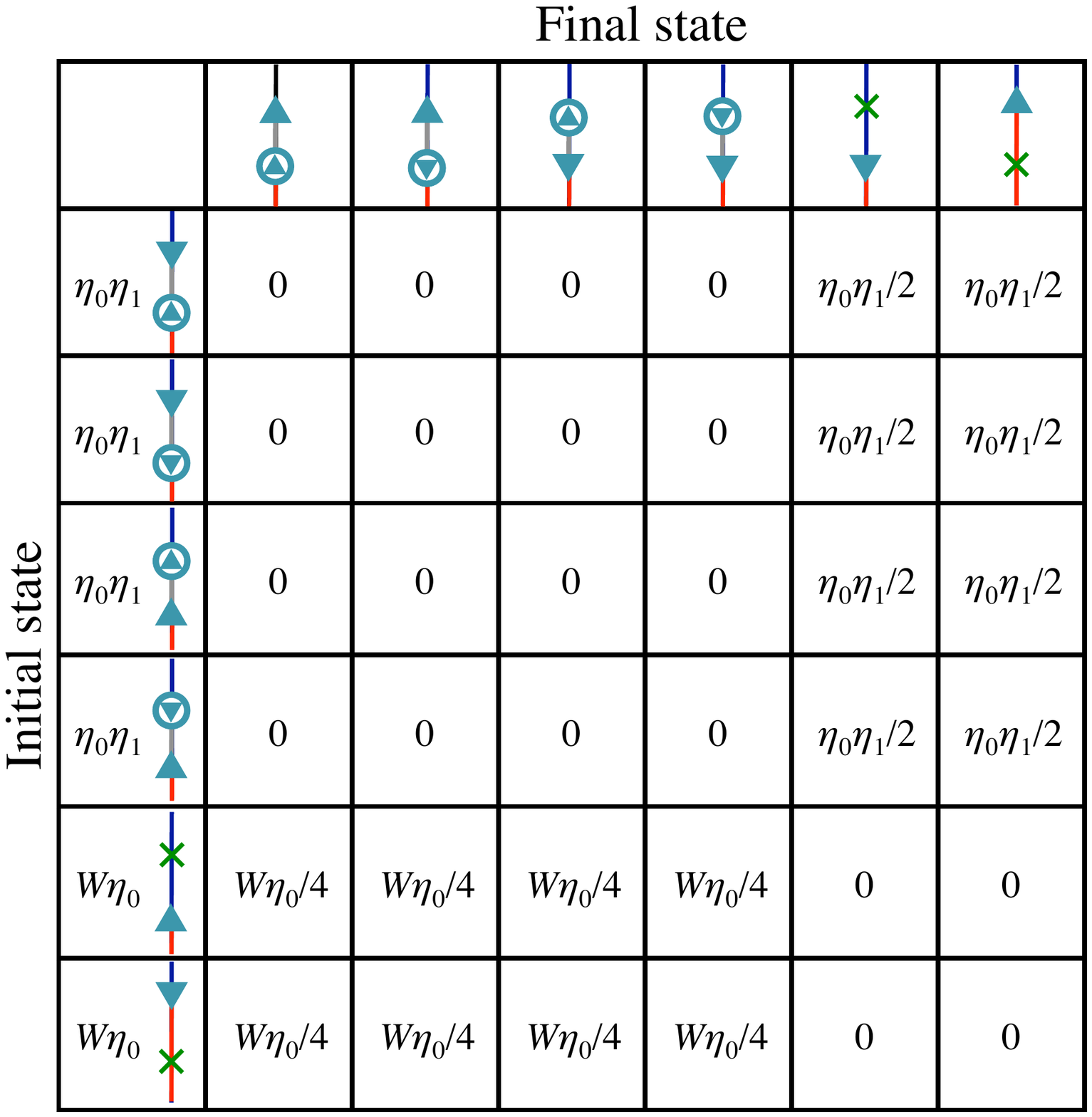}
	\caption{
	  (Color online)
	  Table of $w_{\alpha \beta}$ related to the fission and fusion processes.
	  Green crosses represent the newly introduced single-site vertices for the fission and fusion processes.
	}
\label{fig:app}
\end{figure}

In this appendix, we consider the details of virtual placement of vertices for fission and fusion.
We introduce a single-site vertex for the fission and fusion processes.
The weight of the vertex $W$ is uniform everywhere and independent from the local state of the world-line.
We set $W=2\eta_1$ so that the acceptance probability of fusion is equal to unity.
We start by considering the time-reversal symmetry  of the algorithm 
that is a sufficient condition for satisfying  detailed balance  \cite{kawashima2004}.
Here, we will only consider the time-reversal symmetry condition associated with the fission and fusion processes.
We decompose the weights just before a fission or fusion operation
as $w_{\alpha}=\sum_{\beta}w_{\alpha \beta}$ where $w_{\alpha}$ is the weight of the initial configuration $\alpha$.
By using the partial weights, $w_{\alpha \beta}$, we determine the transition probability from $\alpha$ to $\beta$ as $t_{\alpha\beta}=w_{\alpha\beta}/w_{\alpha}$,
so that the time-reversal symmetry condition, $w_{\alpha}t_{\alpha\beta}=w_{\beta}t_{\beta\alpha}$, is satisfied.
Typically, we take $w_{\alpha\beta}=w_{\beta\alpha}$ for simplicity.
Figure \ref{fig:app} shows an example of $w_{\alpha \beta}$ of the fission and fusion processes.
The table is symmetric when $W=2\eta_1$ and the time-reversal symmetry condition is satisfied.
``Virtual placement of a vertex" means finding the next point along the world line where the head is scattered.
The vertices are generated based on a Poisson distribution with density $W$.
Namely, the time interval to the next vertex is  $-\ln (R)/W$, where  $R$ is random number between 0 and 1: $R \in (0,1]$.

\bibliographystyle{apsrev}
\bibliography{tda}

\end{document}